\documentclass[twocolumn]{article}

\title{THE BLACK HOLE MASS AND MAGNETIC FIELD CORRELATION IN AGN:
 TESTING BY OPTICAL POLARIMETRY}

\author{N.A. Silant'ev, M.Yu. Piotrovich,
 Yu.N. Gnedin\thanks{E-mail: gnedin@gao.spb.ru}\,\, and
 T.M. Natsvlishvili\\ Central Astronomical Observatory at Pulkovo,
 Saint-Petersburg, Russia.}

\begin{document}

\maketitle

\begin{abstract}
We consider the integral light polarization from optically thick
accretion disks. Basic mechanism is the multiple light scattering
on free electrons (Milne's problem) in magnetized atmosphere. The
Faraday rotation of the polarization plane changes both the value
of integral polarization degree $p$ and the position angle $\chi
$. Besides, the characteristic spectra of these values appear. We
are testing the known relation between magnetic field of black
hole at the horizon $B_{BH}$ and its mass $M_{BH}$, and the usual
power-law distribution inside the accretion disk. The formulae for
$p(\lambda)$ and $\chi(\lambda)$ depend on a number of parameters
describing the particular dependence of magnetic field in
accretion disk (the index of power-law distribution, the spin of
the black hole, etc.). Comparison of our theoretical values of $p$
and $\chi $ with observed polarization can help us to choice more
realistic values of parameters if the accretion disk mechanism
gives the main contribution to the observed integral polarization.
The main content is connected with estimation of validity of the
relation between $B_{BH}$ and $M_{BH}$. We found for the AGN NGC
4258 that such procedure does not confirm the mentioned
correlation between magnetic field and mass of black hole.

\end{abstract}

\section{Introduction}

Recently Zhang et al. (2005) used the observed $\nu
L(5100\dot{A})-M_{BH}$ correlation to probe magnetic field of
black holes harbored in AGNs. Their model uses the assumption that
the accretion disk is supplied by the energy injection due to the
magnetic coupling (MC) process, and the gravitational dissipation
due to accretion. They have modeled the relation $\nu
L(5100\dot{A})-M_{BH}$ as a function of the spin parameter
$a/M_{BH}$ and the magnetic field $B_{BH}$. Using the observations
of $\nu L(5100\dot{A})$ for 143 AGNs, they obtain the correlation
between $M_{BH}$ and $B_{BH}$ of the following form:

\begin{equation}
 \log{B_{BH}(G)}=9.26-0.81\log{\frac{M_{BH}}{M_{\odot}}}.
 \label{eq1}
\end{equation}

\noindent According to this formula, for
$M_{BH}/M_{\odot}=10^{11.4}, 10^{9}$, and $10^8$ the magnetic
field $B_{BH}$ is equal to $1$ G, $100$ G, and $630$ G,
respectively. Note that Zhang et al. (2005) have used the theory
of magnetic coupling (see Wang et al. 2002, 2003 and Page \&
Thorne 1974), which supposed that magnetic field is poloidal with
the dominant vertical component to the accretion disk, i.e. in
Eq.(1) is assumed that $B_{BH}\equiv B_{\|}$.

Relation (1) differs significantly from the relationship based on
the standard accretion disk theory of the inner region (see
Shakura and Sunyaev 1973; Novikov and Thorne 1973):

\begin{equation}
 B(G)=10^8\left(\frac{M_{BH}}{M_{\odot}}\right)^{-1/2}
 \left(\frac{3R_g}{r}\right)^{3/4},
 \label{eq2}
\end{equation}

\noindent where $R_g = 2 GM_{BH}/c^2$ is gravitational radius of a
black hole. Eq.~(2) is based on the equipartition between the
radiation pressure and magnetic pressure at the high accretion
rate.

For Shakura-Sunyaev zone (b), where the equipartition between the
gas pressure and magnetic pressure occurs, the magnetic field
strength is determined by another relation:

\[
B(G)=\frac{10^9}{\alpha^{9/20}}
\left(\frac{\dot{M}}{\dot{M}_{Edd}}\right)^{2/5}
\left(\frac{M_{BH}}{M_{\odot}}\right)^{-9/20}\times
\]
\begin{equation}
 \times\left(\frac{3R_g}{r}\right)^{13/10},
 \label{eq3}
\end{equation}

\noindent where $\alpha$ is the Shakura-Sunyaev viscosity
parameter, $\dot{M}$ and $\dot{M}_{Edd}$ are the ordinary and
Eddington accretion rates, respectively.

The typical value of the magnetic field in the optical radiation
region $\approx 10^3R_g$ and for $\alpha = 0.1$ is

\begin{equation}
 B(G)\approx 10^6\left(\frac{\dot{M}}{\dot{M}_{Edd}}\right)^{2/5}
 \left(\frac{M_{BH}}{M_{\odot}}\right)^{-9/20}.
 \label{eq4}
\end{equation}

Narayan and Yi (1995) proposed advection-dominated accretion flow
(ADAF) model. Their model presents the hot advection-disk solution
that does include advection and then dominates the ion energy
transfer. In this case a global magnetic field can be presented as

\[
B(G)=\frac{6.55\cdot 10^8}{\alpha^{1/2}}
\left(\frac{\dot{M}}{\dot{M}_{Edd}}\right)^{1/2}\times
\]
\begin{equation}
 \times \left(\frac{M_{BH}}{M_{\odot}}\right)^{-1/2}
 \left(\frac{R_g}{r}\right)^{5/4}.
 \label{eq5}
\end{equation}

The basic goal of our paper is to calculate the spectrum of linear
polarization that can be used to determine the magnetic field
strength in the region where the optical linear polarization
originates. We calculate the degree and position angle of the
polarization of radiation scattered in accretion flows (standard
accretion disk, ADAF model, etc.). We take into account the
Faraday rotation of the polarization plane (see Dolginov et al.
1995; Gnedin and Silant'ev 1997; Gnedin et al. 2006). The
comparison with the observational data can allow us to test both
the various models of accretion onto a supermassive black hole and
also the correlation between black hole mass and magnetic fields
in AGNs and QSOs. Below we consider the case of relation (1) in
more detail.

\section{Polarization from magnetized accretion disks}

An accretion disk is one of the basic elements of the structure of
accretion flows near supermassive black holes. Since there is no
axial symmetry with respect to the line of sight, the integral,
scattered by free electrons radiation, emerging from an accretion
disk is polarized. The angle of Faraday rotation of a light beam
passing through the distance $l$ is described by the expression
(see, for example, Dolginov et al. 1995; Gnedin and Silant'ev
1997):

\begin{equation}
 \Psi \equiv \frac{1}{2} \delta \tau \cos{\theta} \simeq 0.4
 \left(\frac{\lambda}{1 \mu m}\right)^2 \left(\frac{B}{1 G}\right)
 \tau \cos{\theta},
 \label{eq6}
\end{equation}

\noindent where $\tau $ is Tomson's optical depth of $l$,
$\lambda$ is the wavelength of the radiation and $\theta$ is the
angle between the magnetic field ${\bf B}$ and line of sight ${\bf
n}$.

\begin{equation}
 \delta=\frac{3}{4\pi}\cdot \frac{\lambda}{r_e}\cdot
 \frac{\omega_B}{\omega}\simeq 0.8\lambda^2(\mu m)B(G)
 \label{eq7}
\end{equation}

\noindent is the Faraday dimensionless depolarization parameter.
Here $\omega=2\pi \nu$ is angular frequency, and $\omega_B=eB/m_e
c$ is cyclotron frequency of an electron in a magnetic field,
$r_e=e^2/m_ec^2$ is classic electron radius.

The Faraday rotation of the polarization plane in a magnetized
accretion disk changes both the integral degree of polarization
$p$ and the position angle $\chi$. The spectra of these values
acquire the characteristic forms.

Considering optically thick accretion disks we use the solution of
the Milne problem. Usually one considers the Milne problem
corresponding to sources of thermal radiation being located far
from optically thick plane atmosphere. The exact numerical
solution of this problem for magnetic field ${\bf B}_{\|}$
directed along the normal ${\bf N}$ to the atmosphere are
presented in Agol and Blaes (1996) and in Shternin et al. (2003).

For estimations of magnetic fields in accretion disks we use
simple asymptotical formulae (see Silant'ev 2002 and 2005; Gnedin
et al. 2006, and Silant'ev et al. 2009). These formulae are valid
for arbitrary directions of magnetic field. Firstly we give the
asymptotic formulae for Stokes parameters $I, Q$ and $U$ for
general case of magnetized and absorbing plasma:

\[
I=\frac{F}{2\pi J_1}J(\mu),
\]

\[
Q=-\frac{F}{2\pi J_1}\frac{1-g}{1+g}\frac{(1-\mu^2)(1-k\mu+C)}
{(1-k\mu+C)^2+(1-q)^2\delta^2\cos^2\theta},
\]

\begin{equation}
 U=-\frac{F}{2\pi J_1}\frac{1-g}{1+g}\frac{(1-\mu^2)(1-q)\delta
 \cos\theta} {(1-k\mu+C)^2+(1-q)^2\delta^2\cos^2\theta}.
 \label{eq8}
\end{equation}

\noindent Here, $F$ is the radiation flux emerging from the
surface of accretion disk, $\mu=\cos i$, $i$ is the angle of
inclination of accretion disk, i.e the angle between the outer
normal ${\bf N}$ to the disk and the direction ${\bf n}$ to the
observer. The value $J(\mu)=I(\mu)/I(\mu=0)$ describes the angular
distribution of escaping radiation intensity, $J_1$ is the first
moment of $J(\mu)$, i.e. the integral from $\mu J(\mu)$ over
interval $(0,1)$. Parameter $q$ is the degree of absorption.
Dimensionless parameter $C$ arises in turbulent magnetized plasma
(see Silant'ev 2005), and characterizes the new effect -
additional extinction of parameters $Q$ and $U$ due to incoherent
Faraday rotations in turbulent eddies. Below we present the
apparent formula for this parameter. The value $k$ is the root of
corresponding characteristic equation. For conservative atmosphere
($q=0$) this parameter is equal to zero. The  parameter $g$
characterizes the maximum degree of polarization
$p_{max}=(1-g)/(1+g)$, if we take into account only last
scattering of non-polarized radiation before escape the surface,
neglecting the existence of magnetic field. So. for conservative
atmosphere $g=0.83255$. This corresponds to $p_{max}=9.14\%$.
Remember that the Sobolev - Chandrasekhar value is equal to
11.71\%.

The asymptotic formulae (8) suppose that the depolarization
parameter $\delta$ is large. In this case the terms with
parameters $Q$ and $U$ in full system of transfer equations for
$I, Q$ and $U$ become very small $\sim 1/\delta$, and they are
negligible in the equation for intensity $I$. As a result, the
radiation intensity obeys the separate transfer equation with the
Rayleigh phase function (see, in more detail, Silant'ev 1994). The
first equation in system (8) presents the exact solution of this
equation for the Milne problem. Formulae (8) were obtained in this
way.

The numerical values of $k, g, J_1$ and $J(\mu)$ for various
values $q$ are presented in Silant'ev (2002). So, for example, for
$q=0.1$ we have $k=0.5232, g=0.77129, p_{max}=12.91\%$, and for
$q=0.2$ the  corresponding values are $k=0.70483, g=0.70405$ and
$p_{max}=17.37\%$. It is known, that the polarization degree in
Milne's problem in absorbing atmosphere is greater than in
conservative one. So, for atmospheres with $q=0.1$ and $q=0.2$ the
$p_{max}$-values are equal to 20.4\% and 28.7\%, respectively (see
Silant'ev 1980). In these cases the relative contribution of
parameters $Q$ and $U$ into the polarization itself is greater
than for the case $q=0$ (where this contribution is $\simeq
20\%$). The existence of true light absorption can explain the
observed polarization degrees greater than limiting Sobolev -
Chandrasekhar value 11.71\% for conservative atmosphere. Clearly,
the parameter $q$ depends on wavelength. This gives an additional
mechanism to explain the wavelength dependence of observed
polarization degree $p(\lambda)$ and position angle
$\chi(\lambda)$.

Usually one observes the axially symmetric accretion disks as a
whole. The observed integral Stokes parameters $\langle Q({\bf
n},{\bf B})\rangle $ and $\langle U({\bf n},{\bf B})\rangle $ are
described by the azimuthal averaged formulae

\[
\langle Q({\bf n},{\bf B})\rangle =Q(\mu,0)\,\frac{2}{\pi}(1+C-k\mu)\times
\]
\[
\int_0^{\pi/2}d\Phi
\,\frac{(1+C-k\mu)^2+a^2+b^2\cos^2\Phi}
{[(1+C-k\mu)^2+a^2+b^2\cos^2\Phi]^2-(2ab\cos\Phi)^2},
\]
\begin{equation}
 \langle U({\bf n},{\bf B})\rangle =Q(\mu,
 0)\,a\,\frac{2}{\pi}\times
 \label{eq9}
\end{equation}
\[
\int_0^{\pi/2}d\Phi
\,\frac{(1+C-k\mu)^2+a^2-b^2\cos^2\Phi}
{[(1+C-k\mu)^2+a^2+b^2\cos^2\Phi ]^2-(2ab\cos\Phi )^2}.
\]

\noindent Here $Q(\mu,0)=-F(1-g)(1-\mu^2)/2\pi J_1(1+g)(1-k\mu)$
is the Stokes parameter of scattered non-polarized radiation
$\langle U(\mu,0)\rangle \equiv 0)$ in the classic Milne problem
(see Chandrasekhar 1950). The minus sign denotes that the electric
field oscillations are parallel to accretion disk plane (it means
that the position angle $\chi(\mu, 0)=0$). Further we denote
$p(\mu, 0)\equiv |Q(\mu,0)|$.

For estimations in conservative atmospheres ($q=0$) instead of
$p(\mu, 0)$ one can use $p_{T}(\mu)$ - the known value of
polarization for non-magnetized atmosphere with Thomson (Rayleigh)
scattering, which is presented in Chandrasekhar's book (1950). In
this case we can also use small values of parameter $\delta$,
because the limiting value for $\delta \ll 1$ is $p_{T}(\mu)$.

Dimensionless parameters $a$ and $b$ are connected with the
parallel (along the normal ${\bf N}$ to the disk) ${\bf B}_{\|}$
and perpendicular ${\bf B}_{\bot}$ components of magnetic field
${\bf B}={\bf B}_{\|}+{\bf B}_{\bot}$:

\[
a=(1-q)\delta_{\|}\cos i \simeq 0.8(1-q)\lambda^2(\mu m)B_{\|}(G)\,\mu,
\]
\begin{equation}
 b=(1-q)\delta_{\bot}\sin i \simeq 0.8(1-q)\lambda^2(\mu
 m)B_{\bot}(G)\,\sqrt{1-\mu^2}.
 \label{eq10}
\end{equation}

\noindent Note that the perpendicular magnetic field consists of
two mutually perpendicular components ${\bf B}_{\bot}={\bf
B}_{\rho}+{\bf B}_{\varphi}$, where ${\bf B}_{\rho}$ is radial (
in the plane of a disk) component, and ${\bf B}_{\varphi}$ is the
azimuthal one.

The mentioned above parameter $C$, according to Silant'ev (2005),
is described by formula:

\[
C=(1-q)\tau^{(T)}_1\langle \delta'^2\rangle\,f_B
\]
\begin{equation}
 =0.64(1-q)\tau^{(T)}_1\lambda^4(\mu m)\langle B'^2\rangle f_B/3.
 \label{eq11}
\end{equation}

\noindent Here, $\tau^{(T)}_1$ is the mean Thomson optical length
of turbulent eddies, the values $\delta'$ and $B'$ denote
fluctuating components of corresponding values ($\delta=\delta_0 +
\delta', {\bf B}={\bf B}_0 + {\bf B'})$. In Eqs.(9) and (10) the
values $\delta_{\|}, \delta_{\bot}, {\bf B}_{\|}$ and ${\bf
B}_{\bot}$ are the mean values of corresponding quantities, where,
for brevity, we omitted the subscript "zero". The statistical
averages $\langle\delta'\rangle =0$ and $\langle{\bf B'}\rangle
=0$. The numerical coefficient $f_B\simeq 1$ is connected with the
integral from corresponding correlation function of values $B'$ in
neighboring points of turbulent atmosphere.

The observed degree of the light polarization and the position
angle are derived from parameters (9) by the usual way. For
particular cases of pure normal ($\delta_{\bot}=0$) and pure
perpendicular ($\delta_{\|}=0$) magnetic fields expressions (9)
can be derived analytically. For the first case we have:

\[
p({\bf n},{\bf B})=\frac{p(\mu,0)}
{\sqrt{(1+C-k\mu)^2+(1-q)^2\,\delta^2_{\|}\,\mu^2}},
\]
\begin{equation}
 \tan2\chi=\frac{(1-q)\delta_{\|}\,\mu}{1+C-k\mu}.
 \label{eq12}
\end{equation}

\noindent It is interesting that position angle $\chi$ weakly
depends on parameter $k\mu<1$, which characterises true absorbtion
in the atmosphere. For $C\gg 1$ this term can be omitted. In the
first equation this term can be also omitted, but the nominator
$p(\mu,0)\equiv p(\mu,q,0)$ strictly depends on degree of
absorbtion $q$ (see Silant'ev 2002).

For perpendicular magnetic field are there the formulae:

\[
p({\bf n},{\bf B})=\frac{p(\mu,0)}
{\sqrt{(1+C-k\mu)^2+(1-q)^2\delta^2_{\bot}\, (1-\mu^2)}}
\]
\begin{equation}
 \,\,\,\,\, \chi\equiv 0.
 \label{eq13}
\end{equation}

\noindent For pure perpendicular magnetic field the position angle
$\chi=0$ is due to the axial symmetry of the problem. So, the
electric wave oscillations in this case occur parallel to the
surface of an accretion disk. The case $\chi\neq 0$ can be only
realized if there exists some $B_{\|}$ component.

The degree of linear polarization $p$ and position angle $\chi $
depend on parameters $a$ and $b$ in a fairly complex form. Of
course, the magnitude of polarization decreases with the increase
of $a$ and $b$. The numerical calculations demonstrate that the
relative polarization degree $p({\bf n},{\bf B})/p(\mu,0)$ is
symmetric function of parameters $a$ and $b$. The position angles
$\chi $ do not possess this symmetry. The detailed description of
parameters $p$ and $\chi $, and their wavelength dependence is
presented in Silant'ev et al. (2009). Note that the radial ${\bf
B}_{\rho}$ and azimuthal ${\bf B}_{\varphi}$ components of the
perpendicular magnetic field ${\bf B}_{\bot}$ integrally give the
same degree of polarization.

As it was mentioned, in Eq. (1) we have  $B_{BH}\equiv B_{\|}$. In
this case we can to use more simple expressions following from
Eqs.(12) :

\[
\frac{p(\mu,0)}{p({\bf n}, {\bf
B})}=\frac{(1-q)\delta_{\|}\,\mu}{\sin 2\chi},
\]
\begin{equation}
 1+C-k\mu=\frac{(1-q)\delta_{\|}\,\mu }{\tan 2\chi}.
 \label{eq14}
\end{equation}

\noindent If we know the inclination angle $i$, the observed
polarization degree $p({\bf n},{\bf B})$ and the position angle
$\chi$, then the expressions (14) allow us to know $\delta_{\|}$,
i.e. the value $B_{\|}$, and the value $1+C-k\mu $ for models of
accretion disks with different possible values of degree of
absorbtion $q$. Note, that increase of $p_T(\mu, q)$ occurs more
fast than decrease $(1-q)$ with grow of parameter $q$.

\section{The correlation between $B_{BH}$ and $M_{BH}$, and optical
polarization of AGNs and QSOs}

Using formulae (9), we calculated the polarization degree $p({\bf
n},{\bf B})$ and position angle $\chi({\bf n},{\bf B})$ for the
typical parameters of AGNs considered by Zhang et al. (2005). Our
calculations assume $M_{BH}=10^8\,M_{\odot}$, inclination angle
$i=60^{\circ}$, and effective band width with
$\lambda_{rest}=0.55\mu m$.

The first important step is to derive the scale length
$R_{\lambda}$ defined by the radius in the accretion disk where
the disk temperature matches the wavelength of the monitoring
band. The standard accretion disk (Shakura and Sunyaev 1973) is
characterized by effective temperature profile
$T_e=T_{BH}(R_H/r)^{3/4}$, where $R_H$ is the black hole horizon
radius. There are the series of papers (Koshanek et al. 2006;
Poindexter et al. 2008; Morgan et al. 2008), where the
semi-empirical method of determination of the accretion disk scale
has been developed. The authors used microlensing variability
observed for gravitationally lensed quasars to find the accretion
disk size, and observed (in rest frame) wavelength relation. It is
very important that the scaling is appeared to be consistent with
expectation from the geometrically thin accretion disk model of
Shakura and Sunyaev (1973).

Poindexter et al. (2008) have presented the following scaling
size-wavelength relation:

\[
R_{\lambda}(cm)=0.97\cdot 10^{10}\, \left(\frac{\lambda}{1\mu
m}\right)^{4/3} \times
\]
\begin{equation}
 \left(\frac{M_{BH}}{M_{\odot}}\right)^{2/3}
 \left(\frac{L_{bol}}{\varepsilon L_{Edd}}\right)^{1/3}.
 \label{eq15}
\end{equation}

\noindent Here $R_{\lambda}$ is the radius in accretion disk that
corresponds to the observed effective wavelength $\lambda$,
$L_{bol}$ is the bolometric luminosity, and $L_{Edd}=1.3\cdot
10^{38}(M_{BH}/M_{\odot})$ erg s$^{-1}$ is the Eddington
luminosity, $\epsilon $ is the rest-mass radiation conversion
efficiency. The commonly accepted relation between the $L_{bol}$
of the accretion disk and the accretion rate $\dot {M}$ is
$L_{bol}=\varepsilon \dot {M}c^2$.

First of all we estimate the depolarization parameter $a$ from
Eq.~(10) (the parameter $b$, in this approximation
$B(R_{\lambda})\equiv B_{\|}(R_{\lambda})$, is equal to zero). We
calculate $a$ suggesting the power-law radial dependence of the
magnetic field

\begin{equation}
 B(R_{\lambda})=B_{BH}\left(\frac{R_H}{R_{\lambda}}\right)^n.
 \label{eq16}
\end{equation}

\noindent The radius of the black hole horizon is (Novikov and
Thorne 1973):

\begin{equation}
 R_H(cm)=\frac{GM_{BH}}{c^2}(1+\sqrt{1-a^2_*}),
 \label{eq17}
\end{equation}

\noindent where $a_*$ is the spin of the Kerr black hole.

For our calculations we use the fundamental relation between the
bolometric luminosity and the mass of a supermassive black hole in
AGN (see, for example, McGill et al. 2008; Ziolkowski 2008, Denney
et al. 2009). According to Ziolkowski (2008) we have

\begin{equation}
 M_{BH}(g)=5.71\cdot 10^7 L_{bol}^{0.545\pm 0.036}(44)M_{\odot},
 \label{eq18}
\end{equation}

\noindent where $L_{bol}(44)\equiv L_{bol}/10^{44}$ erg\,s$^{-1}$.
Let us remind also the very important relation $L_{bol}=9\cdot
\lambda L(5100\dot{A})$, where $L(5100\dot{A})$ is differential
luminosity at $\lambda =5100\dot{A}$.

Eqs.~(1), and (15)-(18) allow us to obtain the magnetic field
$B(R_{\lambda})$ for power-law index $n$:

\[
B(R_{\lambda})\equiv B_{\|}(R_{\lambda})=
\]

\begin{equation}
 10^{9.26-2.036n-0.81s +0.055sn}\lambda_{rest}^{-4n/3}f_n(a_*),
 \label{eq19}
\end{equation}

\noindent where $s=\log (M_{BH}/M_{\odot})$ and

\begin{equation}
 f_n(a_*)=\varepsilon^{n/3}(1+\sqrt{1-a^2_*}\,)^n.
 \label{eq20}
\end{equation}

\noindent The analogous formulae can be derived for other models
(see Eqs.~(2)-(5)). Note that in all the formulae magnetic fields
are taken in Gausses, the radii $R_H$ and  $R_{\lambda}$ in cm,
wavelengths in $\mu$m, and mass in grams. To obtain the parameter
$\delta_{\|}$ we substitute the formula (16) to first expression
in Eq.(10).

The coefficient $f_n(a_*)$ is tabulated in Table 1 for a number of
values of parameters $\varepsilon$ and $a_*$. In this Table we
used the known relation between the spin of the Kerr black hole
$a_*$ and the accretion radiation efficiency $\varepsilon$ (see
Tables in  Krolik 2007, and Shapiro 2007).

\begin{table}
\caption[]{\small The values of the coefficient $f_n(a_*)$.}
\setlength{\tabcolsep}{0.15cm}
\centering
\begin{tabular}{l r r r r r r}
\hline
\noalign{\smallskip}
$a_*$; $\varepsilon $ & 0; 0.057 & 0.5; 0.1 & 1; 0.42 & - 0.9; 0.032\\
\noalign{\smallskip}
\hline
\noalign{\smallskip}
$n=0.5$ & 0.88 & 0.93 & 0.86 & 0.68 \\
$n=0.81$& 0.81 & 0.89 & 0.79 & 0.53 \\
$n=1$   & 0.77 & 0.87 & 0.75 & 0.46 \\
$n=5/4$ & 0.72 & 0.84 & 0.70 & 0.37 \\
$n=3/2$ & 0.67 & 0.81 & 0.65 & 0.31 \\
$n=2$   & 0.59 & 0.75 & 0.56 & 0.21 \\
\noalign{\smallskip}
\hline
\end{tabular}
\end{table}

For the same values of parameters $n, \varepsilon$ and $a_*$, as
in Table~1, and the values $\lambda_{rest}=0.55\mu$m,
$M_{BH}=10^8\,M_{\odot}$,\, we calculated first the parameter
$\delta_{\|}$ (see Eqs. (10)). The results are presented in
Table~2. The parameters $\delta_{\|}$ for other wavelengths can be
calculated from $\delta_{\|}(\lambda =0.55\mu$m) by the simple
relation:

\begin{equation}
 \delta_{\|}(\lambda_2)=
 \delta_{\|}(\lambda_1)\left(\frac{\lambda_2}{\lambda_1}\right)
 ^{2-4n/3}.
 \label{eq21}
\end{equation}

The Faraday depolarization is not effective if the parameters
$\delta_{\|}$ is small, $\delta_{\|}<1$. From Eq.(1) we see that
increase of $s=M_{BH}/M_{\odot}$ decreases magnetic field
$B_{\|}$, i.e. the depolarization parameter $\delta_{\|}$ are to
be also decrease. This is confirmed by numerical calculations. So,
for $s=5$ parameter $\delta_{\|}$ lies in the interval (7000 -
40), depending on values $ n, a_{*}$ and $\varepsilon $. For
$s=10$ this parameter acquires the values between 1 and 0.01. The
largest value of parameter $\delta_{\|}$ corresponds to smallest
value of $n$. The values of $\delta_{\|}$, corresponding to
$n=0.5$, are to 100 - 200 times larger than those, corresponding
to $n=2$. It is clearly, why this occurs. Large $n$ give rise to
small values of magnetic fields $B(R_{\lambda})$, as compared with
those corresponding to small values of $n$. Dependence of
$\delta_{\|}$ on parameters $a_{*}$ and $\varepsilon$ is not so
drastic as the $s$ and $n$ -dependencies (see Table 2).

\begin{table}
\caption[]{\small The values of the parameter $\delta_{\|}$.}
\setlength{\tabcolsep}{0.15cm}
\centering
\begin{tabular}{l r r r r r r}
\hline
\noalign{\smallskip}
$a_*$; $\varepsilon $ & 0; 0.057 & 0.5; 0.1 & 1; 0.42 & - 0.9; 0.032\\
\noalign{\smallskip}
\hline
\noalign{\smallskip}
$n=0.5$ &30.337 &32.182 &30.586 &23.347 \\
$n=0.81$&11.463 &12.613 &11.615 & 7.500 \\
$n=1$   & 6.313 & 7.104 & 6.417 & 3.739 \\
$n=5/4$ & 2.880 & 3.338 & 2.939 & 1.496 \\
$n=3/2$ & 1.314 & 1.568 & 1.346 & 0.599 \\
$n=2$   & 0.273 & 0.346 & 0.282 & 0.096 \\
\noalign{\smallskip}
\hline
\end{tabular}
\end{table}

After that we assumed the inclination angle of an accretion disk
$i=60^{\circ}$ and have calculated the parameter $a$  according to
relations (10) for conservative and non-turbulent atmosphere
($q=0, C=0$). Using Eq.~(12), we calculated the polarization
degree $p$ and position angle $\chi$. The results of calculations
are presented in Table~3. Remember that position angle $\chi$ is
equal to zero when electric wave oscillations are parallel to the
accretion disk plane. Remember also that for $i=60^{\circ}$ the
polarization degree $p_T(\mu=0.5)$ is equal to 2.252\% (see
Chandrasekhar 1950). Our calculations show that for relation (1)
the effect of Faraday depolarization acts efficiently in the case
of standard accretion disk for $n<2$. For any distribution of
magnetic field with $n\ge 2$ the depolarization is very low, and
the polarization parameters correspond to the classic limit
$p_T(\mu)$ (Chandrasekhar 1950). This means that there is a direct
relation between the boundary value of $n$ (where depolarization
does not yet occur) and the black hole mass magnitude. So, for
$M_{BH}<10^8M_{\odot}$ the strong depolarization exists for
$n<1.5-2$ (see Table 3).

\begin{table}
\caption[]{\small The values of polarization degree $p\%$, and
position angle $\chi^{\circ}$ for inclination angle of an
accretion disk $i=60^{\circ}$.} \setlength{\tabcolsep}{0.15cm}
\centering
\begin{tabular}{l r r r r r r}
\hline
\noalign{\smallskip}
$a_*$; $\varepsilon $ & 0; 0.057 & 0.5; 0.1 & 1; 0.42 & - 0.9; 0.032\\
\noalign{\smallskip}
\hline
\noalign{\smallskip}
$n=0.5$ & 0.15\%; 43$^{\circ}$ & 0.14; 43  & 0.15; 43 & 0.19; 43 \\
$n=0.81$& 0.39\%; 40$^{\circ}$ & 0.35; 40  & 0.38; 40 & 0.58; 38 \\
$n=1$   & 0.68\%; 36$^{\circ}$ & 0.61; 37  & 0.67; 36 & 1.06; 31 \\
$n=5/4$ & 1.28\%; 28$^{\circ}$ & 1.16; 30  & 1.27; 28 & 1.80; 18 \\
$n=3/2$ & 1.88\%; 17$^{\circ}$ & 1.77; 19  & 1.87; 17 & 2.16; 8 \\
$n=2$   & 2.23\%; 4$^{\circ}$  & 2.22;  5  & 2.23; 4  & 2.25; 1 \\
\noalign{\smallskip}
\hline
\end{tabular}
\end{table}

The technique of estimation of observed polarization from
optically thick magnetized accretion disk, suggested in this
section, uses  relation (12). If the observed polarization differs
strongly from these estimations, it means that formula (12) does
not valid, i.e. the assumptions used by Zhang et al. (2005) are
also invalid. The same technique can be also used for other models
of accretion disk structure. So, the main problem to use this
technique is connected with the observation of polarization degree
$p$ and position angle $\chi$ from AGNs. This is separate, fairly
difficult, problem, especially for estimation of position angle
$\chi $, and its orientation with respect to accretion disk plane.
We do not discuss these problems.

Strictly speaking, our calculations of radiation polarization from
the ring of the radius $R_{\lambda}$ can be used directly if the
gravitational potential $GM_{BH}/R_{\lambda}$ in the observed ring
is much smaller then that near the vicinity of black hole. The
strong gravitational fields near the black holes influences the
Stokes parameters of outgoing radiation when it propagate to a
distant observer. The detailed calculations of this effect have
been made by Connors et al. (1980), and Dovciak et al. (2004).
According to these papers, the results of calculations of the
Stokes parameters $I,Q,$ and $U$ the local (rest) reference frame
in accretion disk are to be integrated along the geodesic paths of
photons escaping from every point of the ring $R_{\lambda}$. This
integration gives rise to the deformation of the spectrum of
radiation (reddening), and to the change in Stokes parameters due
to the rotation of polarization plane. What is important is that
the degree of polarization does not change its value. As a role,
the distance $R_{\lambda}$ for optical wavelengths is far from the
black hole, and our consideration is valid without the
gravitational corrections.

Frequently the constitution of black hole vicinity is more complex
than the existence of accretion disk, one can exists some types of
radiating jets, toroidal clumpy rings etc. The considered
mechanism of light polarization is only one from possible
mechanisms. Besides, usually the observed position angles $\chi$
are not connected directly with the plane of accretion disk. In
this case we are to check that the differences of observed
position angles $\chi_{obs}(\lambda_i)- \chi_{obs}(\lambda_j)$
coincide with the corresponding theoretical differences for
different wavelengths $\lambda_i$ and $\lambda_j$.This procedure
makes the estimation of magnetic field and unknown disk's
inclination angle $i$ more distinct then for the single wavelength
observation.

\subsection{Application of theory to the galaxy NGC 4258}

Now, as an example of using our formulae, consider the Seifert II
galaxy NGC 4258. It is found that this galaxy has accretion disk
(see , for example, Modjaz et al. 2005). The accretion disk has an
almost edge on orientation with inclination angle $i\simeq
83^{\circ}$. In the center there is a black hole with
$M_{BH}\simeq 3.9\cdot 10^7 M_{\odot}$ (see Herrnstein et al.
1999). The polarized continuum radiation from the nucleus of NGC
4258 was observed by Wilkes et al. (1995). We consider here only
observation of continuum polarization at $\lambda \simeq
0.55\mu$m. The polarization degree is equal to $p=0.35\pm 0.01\%$,
and position angle relatively the disk's surface $\chi = 7\pm
1^{\circ}$. For conservative atmosphere the polarization
$p_T(i=83^{\circ})=6.9\%$.

Small polarization degree shows that  $\delta_{\|}$ are to be
great. Indeed, from first Eq.(12), taken for conservative
non-turbulent atmosphere, we obtain $\delta_{\|}\simeq 161.56$,
that corresponds to $B_{\|}(R_{\lambda})=667.6$G and
$\delta_{\|}\mu=19.69$, which gives, according to second Eq.(12),
the value $\chi =43.55^{\circ}$. This angle is far from observed
value $7^{\circ}$. So, it is impossible to explain the
observational data , assuming that accretion disk is the
conservative non-turbulent atmosphere.

Now, we suppose that accretion disk is turbulent ($C\neq 0$) and
conservative medium ($q=0$). In this case Eqs.(14) give rise to
solution: $\delta_{\|}=39.135$, and $C=18.129$. Using Eq.(10) and
(11), we obtain $B_{\|}(R_{\lambda})=161.7$G, and $\tau^{(T)}_1
\langle B'^2\rangle\simeq 929$ (for estimation we take $f_B=1$).
We do not know the characteristic dimension of turbulent eddies.
If we take $\tau^{(T)}_1=0.1$, then we obtain $B'=\sqrt{\langle
B'^2\rangle}=96$G. For $\tau^{(T)}_1=0.5$ the value of $B'$ will
be to 5 times lesser: $B'\simeq 19$G. As far as it goes the
$q$-dependence we know that the absorbing atmosphere has more
great polarization degree $p_T(\mu)$ that the usual conservative
one (see Silant'ev 1980). The term $k\mu$ in Eqs.(1) is small
value (remember that $k<1$ and $\mu=0.122$). So, according to
first Eq.(14), the $\delta_{\|}$ will be greater than that in
conservative case. Neglecting small term $k\mu$ in the second
Eq.(14), we see that the level of magnetic field fluctuations $C$
also increases its value.

It is important that our value $B_{\|}(R_{\lambda})\simeq 161.7$G
we obtained directly from observational data and our asymptotical
formulae (14). We do not use the basic relation (1). The value
$B_{\|}(R_{\lambda})$ from Eq. (19) was derived with the using
this relation. For our case $s=7.591$, $\lambda =0.55\mu$m, and
$B_{\|}(R_{\lambda})=162$G, Eq. (19) acquires the form:

\begin{equation}
 f_n\simeq\frac{B(R_{\lambda},q)}{B(R_{\lambda},0)}\,0.1254\cdot
 10^{1.2723\,n}.
 \label{eq22}
\end{equation}

\noindent For conservative atmosphere and $n=0.5$ we have
$f_{0.5}\simeq 0.54$. For $n=0.81$ one has $f_{0.81}\simeq 1.34$,
for $n=1$ the corresponding value $f_n\simeq 2.35$. Further, with
the increasing $n$ the values $f_n$ increase exponentially. This
behavior is opposite to that presented in Table 1. According to
Table 1 and definition (20), every values of $f_n$ are equal to 1
at $n=0$, and then they diminish monotonically with increasing of
$n$. Our values of $f_n$ in Eq.(22), obtained from observational
polarization and relation (1), are equal to 0.1254 at $n=0$ , and
then they increase very rapidly with increasing $n$.

This means that for NGC 4258 relation (1) distinctly does not
takes place. True absorption increases $\delta_{\|}$ and,
correspondingly, the value $B(R_{\lambda},q)>B(R_{\lambda},0)$.
The absorbing atmosphere increases the $f_n$-values in Eq.(22).

Note that, according to selfconsistent model of Pariev et al.
(2003), parameters $n<1$ correspond to gas density of accretion
dusk which increase with the distance from black hole. It seems
fairly unphysically. For $n>1$, the discrepancy between
theoretical $f_n$ from Table 1, and the values (22) is most large.
It means that it is necessary to take into account the turbulent
magnetic field into accretion disk and its contribution can
radically change the relation between magnetic field strength at
the horizon of a massive black hole and its mass determined by
Zhang et al. (2005).

\section{Wavelength dependence of polarization}

The most important information on the magnetic field distribution
in an accretion disk can be obtained from the wavelength
(frequency) dependence of polarization degree and position angle.
This dependence follows from Eqs.(12) for the case of pure
vertical magnetic field $B_{\|}$, and from Eqs.(13) - for the case
of perpendicular magnetic field $B_{\bot}$. The expressions for
polarization degree in both cases are analogous, but the formulae
for position angles are different. According to second Eq.(13) for
pure perpendicular magnetic field the spectrum of $\chi(\lambda)$
is almost flat. This is can be used, to some extent, for
qualitative characterizing is there great perpendicular magnetic
field, as compared with $B_{\|}$, or not. In general cases we are
to use the numerical calculation of $\langle Q\rangle$ and
$\langle U\rangle $ from Eqs.(9). Because absorption degree $q$
depends on wavelength, it is difficult to speak about wavelength
dependence of $p(\lambda)$ and $\chi(\lambda)$ in general. For
this reason, we restrict ourselves by the case of conservative
atmospheres, where $q=0$ and $k=0$. Such atmospheres are
frequently used in analysis of polarization data.

Remember that parameters $\delta_{\|,\bot}$ are proportional to
$\lambda^2$, and the turbulent parameter $C$ is proportional to
$\lambda^4$. The observed wavelength, according to Wien's
displacement law $\lambda=Const/T_e$, is connected with effective
temperature $T_e$. In most used accretion disk model of Shakura \&
Sunyaev (1973) $T_e\sim r^{-3/4}$. In others models one generally
considers $T_e\sim r^{-s}$, where $r$ is distance from black hole.
According to these relations, we have $R_{\lambda}\sim
\lambda^{1/s}$, where $R_{\lambda}$ is distance corresponding to
emission of $\lambda$-radiation. Using Eq.(16) we conclude that
$B(R_{\lambda})\sim \lambda^{-n/s}$. If $\delta_{\|,\bot}\gg C>1$
the polarization degree spectrum is characterized by the following
expression

\begin{equation}
 p({\bf n},{\bf B})\sim \frac{1}{\delta_{\|,\bot}}\sim
 \frac{1}{\lambda^{2-n/s}}.
 \label{eq23}
\end{equation}

\noindent For most used value $s=3/4$ and $n=5/4$ (see Pariev et
al. 2003), we obtain $p(\lambda)\sim \lambda^{-1/3}$. If one
exists $C\gg \delta_{\|,\bot}>1$, the $p(\lambda)$ - spectrum is
described by another formula:

\begin{equation}
 p({\bf n},{\bf B})\sim\frac{1}{C}\sim \frac{1}{\lambda^4\langle
 B'^2\rangle}.
 \label{eq24}
\end{equation}

\noindent We do not know how $\langle B'^2\rangle$ depends on
$R_{\lambda}$. One can consider two limiting cases. If $\langle
B'^2\rangle =Const$, then $p(\lambda)\sim \lambda^{-4}$. If
$\langle B'^2\rangle \sim B^2$, then $p(\lambda)\sim
\lambda^{-2(2-n/s)}$, i.e. this is square of spectrum (23). In any
case the most sharp dependence, as compared with Eq.(23),
demonstrates that there exists magnetic turbulence.

Now let us say some words about the spectrum of position angle
$\chi$. For $C\gg 1$ the second equation (12) acquires the form

\begin{equation}
 \tan2\chi\simeq \frac{\delta_{\|}\mu}{C}\sim
 \frac{\lambda^{-(2+n/s)}}{\langle B'^2\rangle}.
 \label{eq25}
\end{equation}

\noindent For $\langle B'^2\rangle =Const$ we obtain from Eq.(25)
$\chi(\lambda)\sim \arctan{\lambda^{-(2+n/s)}}$. This gives for
$n=5/4$ and $s=3/4$ the expression $\chi(\lambda)\sim
\arctan{\lambda^{-11/3}}$. For second limiting case $B'\sim
B_{\|}$ we have $\chi(\lambda)\sim \arctan{\lambda^{-(2-n/s)}}$.

Qualitatively the behavior of position angle $\chi$ can be
understood from Eq.(12). If $\delta_{\|}\gg C>1$, we have
$\tan2\chi \gg 1$, and $\chi \to 45^{\circ}$. In this case
$\chi(\lambda )$ does not practically depend on $\lambda$. The
existence of turbulent extinction $C$ diminishes the position
angle. In the limiting case $\chi \ll 1$, we can use the known
relation $\tan 2\chi \simeq 2\chi$. In this case Eq.(25) directly
presents the spectrum of $\chi(\lambda)$. So, for $\langle
B'^2\rangle \simeq Const$, the position angle $\chi \to 0$ very
rapidly, as $\sim \lambda^{-11/3}$ for $n=5/4$ and $s=3/4$. For
$B'\sim B_{\|}$ the position angle tends to zero slowly, as $\sim
\lambda^{-1/3}$ for the same  $n$ and $s$. Note that spectra
$p(\lambda)$ and $\chi(\lambda)$ help us to estimate the
inclination angle $i$, as this was explained in paragraph 3.

\section{Conclusions}

We have showed that the magnetic field strength - the black hole
mass correlation can be probed by the optical polarimetric
observations. The basic idea of this probe is taking into account
Faraday rotation of polarization plane of radiation scattered by
free electrons in optically thick accretion disk. Faraday rotation
gives rise to partial depolarization of outgoing integral
radiation, and to arising of characteristic wavelength spectra of
polarization degree and position angle. Both effects are described
by two depolarization parameters $a$ and $b$ (see Eqs.~(10))
related with normal $B_{\|}$ and tangential $B_{\bot}$ components
of magnetic field inside an accretion disk. The Faraday effect is
stronger if the parameters $a$ and $b$ are greater than unity. In
this case there are the simple relation between the power-law
index $n$ of magnetic field distribution inside the accretion disk
and black hole mass. As a rule. the greater the central black hole
mass $M_{BH}$ the lower the power-law index.

For massive $M_{BH}\approx (10^8 - 10^9)M_{\odot}$ one should
expect the polarization degree corresponding to classic electron
scattering without noticeable Faraday depolarization. Only for
extremely smooth ($n\le 1$) magnetic field distribution one can
expect a display of depolarization effects. For low massive
$M_{BH}\approx (10^5 - 10^6)\,M_{\odot}$ black holes the
depolarization effects can be displayed even for steep
distribution of magnetic field $n\ge 1.5-2$.

As an example of application of presented theory we considered the
AGN NGC 4258, where is found the accretion disk, which certainly
plays the main role in polarization emission in continuum. It was
found that for this source the basic relation (1) between magnetic
field and the mass of black hole does not takes place. This
negative result gives rise to basic question for what sources the
correlation relation (1) really exists and why.

\section*{Acknowledgements}

This research was supported by the Grant of President of Russian
Federation ``The Basic Scientific Schools", NS-61110.2008.2, by
FEBR (Project No. 07-02-00535a), Program of Prezidium of RAS , the
Program of the Department of Physical Sciences of RAS.
M.Yu.Piotrovich acknowledges the Council of Grants of President of
Russian Federation for Young Scientists, grant No. 4101.2008.2.

\end{document}